\documentclass[epj,final]{svjour}% referee final
\usepackage{graphics}
\begin{document}
\title{Succesful renormalization of a QCD-inspired Hamiltonian}
\author{Hans-Christian Pauli}
\institute{Max-Planck-Institut f\"ur Kernphysik, D-69029 Heidelberg,
           \email{pauli@mpi-hd.mpg.de}}
\date{18 September 2003}% 
\abstract{%
    The long standing problem of non perturbative renormalization 
    of a gauge field theoretical Hamiltonian is addressed and
    explicitly carried out within an (effective) light-cone 
    Hamiltonian approach to QCD. 
    The procedure is in line with the conventional ideas: 
    The Hamiltonian is first regulated by suitable cut-off functions,
    and subsequently renormalized by suitable counter terms to make
    it cut-off independent. Emphasized is the considerable freedom
    in the cut-off function which eventually can modify the Coulomb 
    potential of two charges at sufficiently small distances. 
    The approach provides new physical insight into nature of 
    gauge theory and the potential energy of QCD and QED near the origin.
    The so obtained formalism is applied to physical mesons with
    a different flavor of quark and anti-quark.
    The excitation spectrum of the $\rho$-meson with its
    excellent agreement between theory and experiment
    is discussed as a pedagogical example.
\PACS{{11.10.Ef}%{Lagrangian and Hamiltonian approach}
 \and {12.38.Aw}%{General properties of QCD} 
 \and {12.38.Lg}%{Other non perturbative calculations}   
 \and {12.39.-x}%{Phenomenological quark models}
   {}} 
} 
\maketitle
%
%\tableofcontents
%
\section{Introduction} 
\label{sec:1}
When starting in 1984 with 
Discretized Light-Cone Quantization (DLCQ) \cite{PauBro85a} 
and with a
revival of Dirac's Hamiltonian front form dynamics \cite{dir49}, 
all challenges of a gauge field Hamiltonian theory 
were essentially open questions, particularly 
   the non perturbative bound state problem,
%  Lorenz and gauge invariance, 
   the many-body aspects,   
   regularization,  
   renormalization,  
   confinement, 
   chirality,
   vacuum structure and condensates,
just to name a few. 
The step from the gauge field QCD Lagrangian
down to a non relativistic Schr\"odinger equation 
was completely mysterious.
Now we know better \cite{BroPauPin98}. 
We have learned how to partition the problem and how 
to shape our thinking in four major steps:
\begin{eqnarray}
  \begin{array}{crcl}
                   & &\mathcal{L}_\mathrm{QCD}& ,                           \\
   \hookrightarrow & H_\mathrm{LC} \vert\Psi\rangle &=& M^2\vert\Psi\rangle,\\
   \hookrightarrow & H_\mathrm{eLC} \vert \Psi _{q\bar q}\rangle &=& 
                                           M^2 \vert \Psi _{q\bar q}\rangle,\\
   \hookrightarrow & H_\mathrm{eff}\vert\varphi\rangle &=& 
                                                       E\vert\varphi\rangle,\\
   \hookrightarrow & %\displaystyle
   \big(\frac{\vec p^{\,2}}{2m_r}+V(\vec r)\big)\psi(\vec r) 
                                                           &=& E\psi(\vec r).
  \end{array}
\label{eq:1}\end{eqnarray}
We have understood, for example, that 
the chiral phase transition, in which the quarks are supposed
to get their mass, is not the major challenge.
The challenge is to understand what happens \emph{after} 
the phase transition, at zero temperature.
The challenge is to understand the spectrum of physical hadrons
and to get the corresponding eigenfunctions, the light cone wave functions. 

%
%%%%%%%%%%%%%%%%%%%%%%%%%%%%%%%%%%%%%%%%%%%%%%%%%%%%%%%%%%%%%% beg figure
\begin{figure*}\sidecaption
   \resizebox{0.24\textwidth}{!}{\includegraphics{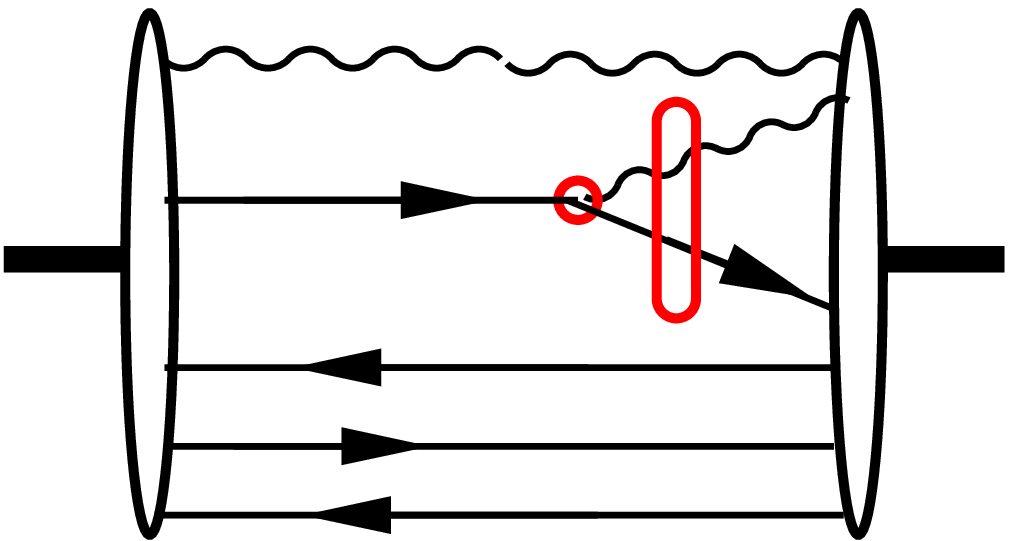}
}  \resizebox{0.38\textwidth}{!}{\includegraphics{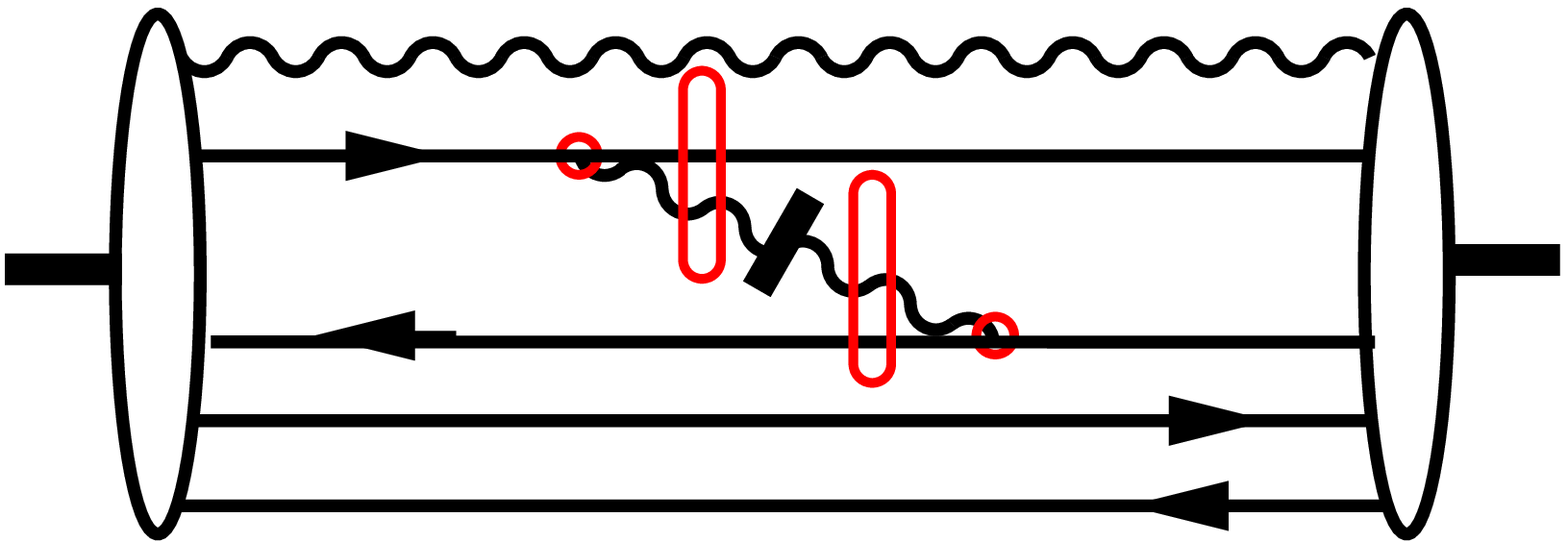}
}\caption{ 
   Regularization of the interaction by vertex regularization. 
   In a matrix element, as illustrated on the left for a vertex,
   a quark changes its four-momentum from $k_1$ to $k_2$, 
   \textit{i.e.} $Q^2= -(k_1-k_2)^2$. 
   The vertex interaction is regulated by multiplying with a 
   form factor $F(Q^2)$, as indicated by the circle.~--
   Instantaneous interactions are treated correspondingly,
   as illustration on the right for a seagull.
}\label{fig:reg}\end{figure*}
%%%%%%%%%%%%%%%%%%%%%%%%%%%%%%%%%%%%%%%%%%%%%%%%%%%%%%%%%%%%%% end figure
%
The light-cone wave functions $\Psi$ for a hadron with mass $M$
encode all possible quark and gluon
momentum, helicity and flavor correlations and, in principle, 
are obtained by diagonalizing the QCD light-cone Hamiltonian
$H_\mathrm{LC} = P^+P^- - P^2_{\!\perp}$, 
where $P^{\pm} = P^0 \pm P^z$, 
\begin{eqnarray} 
   H_\mathrm{LC}\vert\Psi_i\rangle = M_i^2 \vert\Psi_i\rangle
\label{aeq:2}\end{eqnarray} 
in a complete basis of Fock states with increasing complexity. 
For example, the positive pion has the Fock expansion:
\begin{eqnarray*}
   \vert\Psi_{\pi^+}\rangle 
   &=& \sum _n \langle n\vert\pi^+\rangle \vert n\rangle  
\\ &=&   
   \Psi^{(\Lambda)}_{u\bar d/\pi}(x_i,\vec{k}_{\!\perp i}) 
   \vert u\bar d\rangle +
   \Psi^{(\Lambda)}_{u\bar dg/\pi}(x_i,\vec{k}_{\!\perp i})
   \vert u\bar dg\rangle  + \dots 
\;,\end{eqnarray*}
representing the expansion of the exact QCD eigenstate at scale $\Lambda$ 
in terms of non-interacting quarks and gluons.
The $i_n$ particles in a Fock state ($n$) have 
longitudinal light-cone momentum fractions $x_i$
and relative transverse momenta $\vec{k}_{\!\perp i}$, with
\begin{eqnarray*}
   x_i = \frac {k_i^+}{P^+} = \frac {k_i^0+k_i^z}{P^0+P^z}\;, \quad
   \sum_{i=1}^{i_n} x_i =1\;, \quad
   \sum_{i=1}^{i_n} \vec{k}_{\!\perp i} = \vec 0_{\!\perp} 
\;.\end{eqnarray*}
The form of $\Psi_{n/H}(x_i,\vec{k}_{\!\perp i})$ is invariant under
longitudinal and transverse boosts; i.e., the light-cone wave functions 
expressed in the relative coordinates $x_i$ and $k_{\!\perp i}$ are 
independent of the total momentum ($P^+$, $\vec P_{\!\perp}$) of the hadron. 
The first term in the expansion is referred to as the valence Fock state,
as it relates to the hadronic description in the constituent quark model.
The higher terms are related to the sea components of the hadronic 
structure.
It has been shown that the rest of the light-cone wave function 
is determined once the valence Fock state is known \cite{Mue94,Pau99b},
with explicit expressions given in \cite{Pau99b}.

The key issue is to overcome the problem of any gauge theory,
that the unregulated theory exposes logarithmic singularities.
The problem of regularization and renormalization has been solved in the 
perturbative context of scattering theory, but not in the non perturbative 
context of a Hamiltonian. It is addressed to in the first two sections and 
applied in the remainder of this paper. 

\section{Regularization}
\label{sec:2}
Canonical field theory with the conventional QCD Lagrangian 
allows to derive the components of the total canonical 
four-momentum $P^\mu$.
Its front form version \cite{BroPauPin98} rests on two assumptions, 
the light cone gauge $A^+=0$ \cite{LepBro80} and the
suppression of all zero modes \cite{BroPauPin98,Kal95}.
The front form vacuum is then trivial.

I find it helpful to discuss the problem in terms of 
DLCQ \cite{PauBro85a,BroPauPin98}.
In the back of my mind I visualize an explicit finite dimensional 
matrix representation of the Light-Cone Hamiltonian 
as it occurs for finite harmonic resolution. 
Such one is schematically displayed in Fig.~2 of \cite{BroPauPin98}. 
All of its matrix elements are
finite for any finite $x$ and $k_{\!\perp}$. 

The problem arises for ever increasing harmonic resolution,
on the way to the continuum limit: 
The numerical eigenvalues 
are numerically unstable and diverge logarithmically \cite{KraPauWoe92,TriPau00},
contrary to the calculations in 1+1 dimension \cite{PauBro85a};
see also actual DLCQ calculations in 3+1 by Hiller \cite{Hil00}.

The reason is inherent to Dirac's   
relativistic vertex interaction 
$\langle k_1,h_1\vert V \vert k_2,h_2;k_3,h_3\rangle$, 
in which some particle `1' is scattered into two particles `2' and `3'
with their respective four-momenta $k$ and helicities $h$,
see Fig.~\ref{fig:reg}.
The matrix element for bremsstrahlung, for example,
is proportional to $k _{\!\perp}$,
$  \langle k_1,\uparrow\vert V \vert k_2,\uparrow;
   k_1,\uparrow\rangle 
   \propto \vert\vec k _{\!\perp}\vert$, 
see Table~9 in \cite{BroPauPin98},
when the quark maintains its helicity 
while irradiating a gluon with four-momentum 
$k_3^\mu=(xP^+,\vec k _{\!\perp},k_3^-)$. 
Singularities arise typically when squares of such matrix elements 
are integrated over all $\vec k _{\!\perp}$ as in the  
integrations of perturbation theory.

The singularities are avoided \textit{a priori}
by \emph{vertex regularization}, by multiplying each (typically off-diagonal) 
matrix element with a regulating \emph{form factor} $F$:
\begin{eqnarray}
   &&
   \langle k_1,h_1\vert V \vert k_2,h_2;k_3,h_3\rangle \;\Longrightarrow
\nonumber\\ 
   && 
   \langle k_1,h_1\vert V \vert k_2,h_2;k_3,h_3\rangle \; F(Q) 
\;.\label{eq:2}\end{eqnarray}
It took several years to realize that it is the Feynman 
four-momentum transfer across a vertex, $Q^2 = -(k_1-k_2)^2$, 
which governs any effective interaction. 
The minimal requirement for such a form factor is 
\begin{equation} 
    F(Q;\Lambda)=\left\{
    \begin{array}{ll}
      1\;,&\mbox{ for } Q^2 \rightarrow0     \;,\\  
      0\;,&\mbox{ for } Q^2 \rightarrow\infty\;.
    \end{array}\right.
\label{eq:3}\end{equation}
The job would be done by a step function, 
$F(Q)=\Theta(Q^2-\Lambda^2)$.
The limit $\Lambda\rightarrow0$ suppresses the interaction
all together, the limit $\Lambda\rightarrow\infty$
restores the interaction and its problems. 
Any finite value of $\Lambda^2$ restricts $Q^2$ to be finite 
and eliminates the singularities.
But the sharp cut-off generates problems in an other corner of the theory  
and $F(Q)$ must be an analytic function of $Q$, as to be seen below. 

\emph{Vertex regularization} takes thus care 
of the ultraviolet divergences.
The (light-cone) infrared singularities are taken care of 
as usual by a kinematical gluon mass. 

As usual, regularization is not unique and many ways can do that. 
Dimensional regularization, for example, is not applicable
in a matrix approach which is stuck with the precisely 3+1 dimensions of
the physical world.
\emph{Vertex regularization} should be confronted with the 
\emph{Fock space regularization} 
of Lepage and Brodsky \cite{LepBro80}, see also \cite{BroPauPin98}, 
which has blocked the renormalization aspects for many years.
It also should be confronted with \cite{wil89} 
and \cite{WilWalHar94}.
After applauding the light-cone approach \cite{wil89},
Wilson and collaborators \cite{WilWalHar94} have attempted to
base their considerations almost entirely 
on a renormalization group analysis,
but no concrete technology has emerged thus far.

\section{Renormalization}
\label{sec:3}
The non perturbative renormalization of the Hamiltonian 
was stuck for many years by the fact that the coupling constant $g$ 
and the regulator function $F(Q)$
multiply each other in Eq.(\ref{eq:2}).
It was always clear that one may add \emph{non local counter terms} 
\cite{WilWalHar94}, but is was not clear how they could be constructed.
Progress has come from recent work on a particular model 
\cite{FrewerFrePau02}, 
which did allow to formulate a paradigmatic example 
in modern renormalization theory.

Here is the general but abstract procedure.

Suppose to have solved Eq.(\ref{aeq:2}) for a fixed value of 
the 7 `bare' parameters in the Lagrangian,
for the coupling constant $g=g_0$ and the 6 flavor quark masses $m_f=(m_f)_0$,
and for a fixed value of exterior cut-off scale $\Lambda=\Lambda_0$.
Suppose further that these 7+1 parameters are chosen such,
that the  calculated $M^2_i$ agree with the corresponding experimental values.
Next, suppose to change the cut-off  by a small amount $\delta\Lambda$.
Every calculated eigenvalue will then change by $\delta M^2_i$.
Renormalization theory is then the attempt to reformulate the
Hamiltonian, such, that all changes $\delta M^2_i$ vanish identically.

The fundamental renormalization group equation is therefore:
\begin{equation}
   \left. d M^2_i \right\vert_{0}=
   \left. d M^2_i \right\vert_{g=g_0,m_f=m_{f_0},\Lambda=\Lambda_0}=0 
\;,\label{eq:4}\end{equation}
\emph{for all eigenstates} $i$. 
Equivalently one requires that 
\emph{the Hamiltonian is stationary} with respect to small $\delta \Lambda$:
\begin{eqnarray}
   \left.\phantom{M^2_i}\delta  H_\mathrm{LC}\right\vert_{0} = 0
\;.\end{eqnarray}
Hence forward reference to ($g_0,m_{f_0},\Lambda_0$), 
to the `renormalization point', will be suppressed.

The Hamiltonian can be made stationary by making $g$ and the $m_f$ functions
of $\Lambda$, by introducing \emph{physical} coupling constants and masses,
$\overline g$ and $\overline m_f$, respectively, which themselves are functions
of the bare $g$ and $m_f$, and which are functionals of the regulator 
$\overline F=F$.
The variation of $H_\mathrm{LC}$ reads then 
\begin{eqnarray*}
   \delta H_\mathrm{LC} = 
   \delta \overline g \frac{\delta H_\mathrm{LC}}{\delta\overline\alpha} + \sum_f
   \delta \overline m_f \frac{\delta H_\mathrm{LC}}{\delta\overline m _f} +
   \delta \overline F \frac{\delta H_\mathrm{LC}}{\delta\overline F} = 0 
\;,\end{eqnarray*}
with the familiar variational derivatives. 
However, since $\overline g$ and $\overline m_f$ are themselves 
functionals of $\overline F$, this reduces to
\begin{eqnarray*}
   \delta H_\mathrm{LC} = 
   \delta \overline F \frac{\delta H_\mathrm{LC}}{\delta\overline F} = 0 
\;.\end{eqnarray*}
Eq.(\ref{eq:4}) as the fundamental equation of renormalization theory 
is then replaced by
\begin{eqnarray}
   \delta \overline F = \delta \Lambda
   \frac{\partial\overline F }{\partial\Lambda}= 0 
\;,\label{eq:6a}\end{eqnarray}
since the variational derivative of the Hamiltionian with respect 
to the regulator is unlikely to vanish. 

It can be solved by counter term technology, as follows. 
A counter term is added to the Hamiltonian, whose interaction has exactly
the same structure except that the regulator $F(Q)$ is replaced
by $C(Q)$. This defines 
\begin{eqnarray}
    \overline F(Q,\Lambda) =
    F(Q,\Lambda) + C(Q,\Lambda)
\;,\label{eq:7}\end{eqnarray} 
subject to the constraint that the counter term vanishes at the 
renormalization point,
\begin{eqnarray}
    \left.\phantom{\frac{d\overline R}{d\Lambda}}
    C(Q,\Lambda)\right|_{\Lambda=\Lambda_0}=0
\;.\label{eq:8}\end{eqnarray} 
The fundamental equation (\ref{eq:6a}) defines then a differential equation
\begin{equation}
    \frac{dC(Q;\Lambda)}{d\Lambda} = - \frac{dF(Q;\Lambda)}{d\Lambda} 
\;,\end{equation} 
which, in its  integral form, includes the initial condition 
\begin{equation}
    C(Q,\Lambda) = - \int\limits_{\Lambda_0}^{\Lambda}
    \!\!ds\ \frac{dF(Q,s)}{ds} = 
    F(Q,\Lambda_0) - F(Q,\Lambda) 
\;.\label{eq:9}\end{equation}
The renormalized regulator function, $\overline F=F+C$,
\begin{equation}
    \overline F(Q,\Lambda) = F(Q,\Lambda_0)
\;,\label{eq:10}\end{equation}
is \emph{manifestly independent of $\Lambda$}.
By construction, the value of $\Lambda_0$ is determined by experiment.

One should emphasize an important point:
In deriving Eq.(\ref{eq:10}), use was made of assuming the regulator 
function has well defined derivatives with respect to $\Lambda$.
The theta function of the sharp cut-off, however, 
is a distribution with only ill defined derivatives.

This raises an other important point: If $F(Q,\Lambda)$ is a function
of $Q/\Lambda$ other than a theta function, one must specify  
how the function approaches the limiting values of Eq.(\ref{eq:3}). 
The case of the `soft' regulator
\begin{eqnarray}
   F(Q,\Lambda) = \frac{\Lambda^2}{\Lambda^2+Q^2}
\label{eq:12}\end{eqnarray}
is only a very special example. In a more general approach 
the soft regulator plays the role of a generating function
\begin{eqnarray}
   F(Q,\Lambda) = \left[1+\sum_{n=1}^{N}
   (-1)^n s_n \Lambda^n\frac{\partial^n}{\partial\Lambda^n}\right]
   \frac{\Lambda^2}{\Lambda^2+Q^2} 
\;.\label{eq:13}\end{eqnarray}
The partials $\Lambda^n\,\partial^n/\partial\Lambda^n$ are 
dimensionless and independent of a change in $\Lambda$.
The arbitrarily many coefficients $s_1,\dots,s_N$ are 
renormalization group invariants and, as such, subject to be
determined by experiment.

\section{The effective (light-cone) Hamiltonian}
\label{sec:4}

In a field theory, one is confronted with a many-body problem
of the worst kind: Not even the particle number is conserved.
For to formulate effective Hamiltonians more systematically,
a novel many-body technique had to be developed,
the \emph{method of iterated resolvents} \cite{Pau99b,Pau98},
whose details are not important here.
 
Important is that 
the \emph{effective light-cone Hamiltonian} $H_\mathrm{eLC}$
has the same eigenvalue as 
the \emph{full light-cone Hamiltonian} $H_\mathrm{LC}$ 
and that it generates the bound state wave function of valence quarks
by an one-body integral equation
in  ($x,\vec k_{\!\perp}$):
\begin{eqnarray} 
\lefteqn{\hspace{-2em}
    M^2\psi_{h_1h_2}(x,\vec k_{\!\perp}) = 
    \left[ 
    \frac{\overline m^2_{1} + \vec k_{\!\perp}^{\,2}}{x} +
    \frac{\overline m^2_{2} + \vec k_{\!\perp}^{\,2}}{1-x}  
    \right]
    \psi_{h_1h_2}(x,\vec k_{\!\perp})  
}\nonumber\\ 
    &-& {1\over 3\pi^2}
    \sum _{ h_q^\prime,h_{\bar q}^\prime}
    \!\int\!
    \frac{dx^\prime d^2 \vec k_{\!\perp}^\prime}
    {\sqrt{ x(1-x) x'(1-x')}}
    \;\psi_{h_1'h_2'}(x',\vec k_{\!\perp}')
\nonumber\\  
    &\times&
    F(Q_{q}) F(Q_{\bar q})
    \left(\frac{\overline\alpha(Q_{q})}{2Q_{q}^2} +
    \frac{\overline\alpha(Q_{\bar q})}{2Q_{\bar q}^2}\right)
\nonumber\\  
    &\times&
    \left[\overline u(k_1,h_1)\gamma^\mu u(k_1',h_2')\right]
    \left[\overline v(k_2',h_2')\gamma_\mu v(k_2,h_2)\right] 
\;.\label{eq:14}\end {eqnarray}
One has achieved step 2 of Eq.(\ref{eq:1}):
$ H_\mathrm{eLC} \vert \Psi _{q\bar q}\rangle = 
  M^2 \vert \Psi _{q\bar q}\rangle $.
Here, $M ^2$ is the eigenvalue of the invariant-mass squared. 
The associated eigenfunction $\psi_{h_1h_2}(x,\vec k_{\!\perp})$ 
is the probability amplitude 
$\langle x,\vec k_{\!\perp},h_{1};1-x,-\vec k_{\!\perp},h_{2}
\vert\Psi_{q\bar q}\rangle$ 
for finding the quark with momentum fraction $x$, 
transversal momentum $\vec k_{\!\perp}$ and helicity $h_{1}$,
and correspondingly the anti-quark.
Expressions for 
the (effective) quark masses $\overline m _1$  and $\overline m _2$ 
and the (effective) coupling function $\overline\alpha(Q)$ are given 
in \cite{Pau98}.
$Q_q$ and $Q_{\bar q}$ are the Feynman momentum transfers 
of quark and anti-quark, respectively,
and $u(k_1,h_1)$ and $v(k_2,h_2)$ are their 
Dirac spinors in Lepage Brodsky convention \cite{LepBro80}, 
given explicitly in \cite{BroPauPin98}.
They arrange themselves in the Lorenz scalar spinor matrix 
\begin{eqnarray*}
   \langle h_1,h_2\vert S\vert h_1',h_2'\rangle &=&
   \left[\overline u(k_1,h_1)  \gamma^\mu u(k_1',h_1')\right] 
\\ &\times&
   \left[\overline v(k_2',h_2')\gamma_\mu v(k_2,h_2)\right]
\,\end{eqnarray*}
which is a rather complicated (matrix) function of its
six arguments $x,x',\vec k_{\!\perp},\vec k_{\!\perp}'$,
as tabulated in \cite{Pau00c}.   
Finally, the form factors $F(Q)$ restrict the range of 
integration and regulate the interaction.
Note that the equation is fully relativistic and covariant.
   
It should be emphasized that Eq.(\ref{eq:14}) is valid only for 
quark and anti-quark having different flavors \cite{Pau99b,Pau98}.
The additional annihilation term for identical flavors is omitted.
At present, it is investigated by \cite{Kra04}.
It should also be emphasized that the same structure was obtained
with a completely different method, with 
Wegner's Hamiltonian flow equations \cite{Wegner00}.
In \cite{Wegner00} is also shown why the concept 
of a `mean momentum transfer',
$Q^2=\frac12\left(Q_{q}^2+Q_{\bar q}^2\right)$ 
is a meaningful simplification. It allows to replace
Eq.(\ref{eq:14}) by 
\begin{eqnarray} 
\lefteqn{\hspace{-2em}
    M^2\psi_{h_1h_2}(x,\vec k_{\!\perp}) = 
    \left[ 
    \frac{\overline m^2_{1} + \vec k_{\!\perp}^{\,2}}{x} +
    \frac{\overline m^2_{2} + \vec k_{\!\perp}^{\,2}}{1-x}  
    \right]
    \psi_{h_1h_2}(x,\vec k_{\!\perp})  
}\nonumber\\ 
    &-& {1\over 3\pi^2}
    \sum _{ h_q^\prime,h_{\bar q}^\prime}
    \!\int\!\displaystyle 
    \frac{dx^\prime d^2 \vec k_{\!\perp}^\prime
    \;\psi_{h_1'h_2'}(x',\vec k_{\!\perp}')}
    {\sqrt{ x(1-x) x'(1-x')}}
    \frac{\overline\alpha(Q)}{Q^2} \overline  R(Q) 
\nonumber\\  
    &\times&
    \left[\overline u(k_1,h_1)\gamma^\mu u(k_1',h_2')\right]
    \left[\overline v(k_2',h_2')\gamma_\mu v(k_2,h_2)\right] 
\;.\label{eq:15}\end {eqnarray}
The form factors $F(Q)$ have made their way into the regulator function
$\overline R(Q)=F^2(Q)$. Krautg\"artner \textit{et al} \cite{KraPauWoe92}
and Trittmann \textit{et al} \cite{TriPau00} have shown 
how to solve numerically such an equation with a high precision. 
But since the numerical effort is so considerable, 
it is reasonable to work first with (over-)simplified models, 
as specified next.

\textbf{The Singlet-Triplet model}.  
Quarks are at relative rest
when $\vec k _{\!\perp}= 0$ and $ x = \overline x$,
with $ \overline x \equiv \overline m_1/(\overline m_1+\overline m_2)$.
An inspection of Eq.(33) in \cite{Pau00c} reveals that   
for very small deviations from the equilibrium values, 
the spinor matrix $\langle h_1,h_2\vert S\vert h_1',h_2'\rangle$
is proportional to the unit matrix, 
\begin{eqnarray}
   \langle h_1,h_2\vert S\vert h_1'h_2'\rangle  
   &\simeq & 4 \overline m_1 \overline m_2 
   \;\delta_{h_1,h_1'}
   \;\delta_{h_2,h_2'}
\;.\end{eqnarray}
For very large deviations, particularly for
$\vec k_{\!\perp}^{\prime\,2} \gg \vec k_{\!\perp} ^{\,2}$,
holds  
\begin{eqnarray}
   Q ^2 \simeq\vec k_{\!\perp}^{\prime\,2}
   \;,\quad\mbox{and}\quad
   \langle\uparrow\downarrow\vert S\vert\uparrow\downarrow\rangle 
   \simeq 2\vec k_{\!\perp}^{\prime\,2}
\;.\label{eq:17}\end{eqnarray}
The \emph{Singlet-Triplet (ST) model} combines these aspects:
\begin{eqnarray} 
    \langle h_1,h_2\vert S\vert h_1',h_2'\rangle 
    &=&
    \delta_{h_1,h_1'}\;\delta_{h_2,h_2'}\;
    \langle h_1,h_2\vert S\vert h_1,h_2\rangle
\;,\label{eq:18}\\
    \frac{\langle h_1,h_2\vert S\vert h_1,h_2\rangle}{Q^2}
    &=&
    \left\{
    \begin{array}{ll}
    \frac{4\overline m_1\overline m_2}{Q^2}+2, 
      &\mbox{ for $h_1 = - h_2 $,}\\
    \frac{4\overline m_1\overline m_2}{Q^2},\phantom{+2} 
      &\mbox{ for $h_1 = \phantom{-} h_2 $.}\\
    \end{array}\right.
\label{eq:19}\end{eqnarray} 
For anti parallel helicities $h_1 = - h_2 $ (singlets)
the model interpolates between two extremes:
For small momentum transfer $Q$, the `2' in Eq.(\ref{eq:17}) is unimportant and 
the Coulomb aspects of the first term prevail.
For large $Q$, the Coulomb aspects are
unimportant and the hyperfine interaction is dominant.
The `2' carries the \emph{singlet triplet mass difference}:
Its value is understood by 
$g_s\left(\frac{1}{4}-\left(-\frac{3}{4}\right)\right)=g_s$,
with the spin-g factor $g_s=2$.
For parallel helicities $h_1 = h_2 $ (triplets) the model 
reduces to the Coulomb kernel.
The model over emphasizes many aspects but its
simplicity has proven useful for fast and analytical calculations. 
Most importantly, the model allows to drop the helicity summations
which technically simplifies the problem enormously.
A more detailed investigation of the spinor matrix 
can be found in \cite{KrassPau02}.

The model can not be justified in the sense of an approximation,
but it emphasizes the point that the `2', or any other constant 
in the kernel of an integral equation, leads to numerically
undefined equations and thus singularities.
Replacing the function $\overline \alpha(Q)$ by the strong coupling constant
$\alpha_s=g^2/4\pi$ completes the model assumptions.
Hence forward, the overline bars for the effective quantitites will be
suppressed.
\section{The potential energy}
\label{sec:5}
It is possible to subtract a c-number from $H_\mathrm{eLC}$
and to define an effective Hamiltonian $H_\mathrm{eff}$ implicitly by
\begin{eqnarray}
   H_\mathrm{eLC}\equiv \left(m_1+m_2\right)^2 +
   2\left(m_1+m_2\right)H_\mathrm{eff}
\;.\end{eqnarray}
Its eigenvalues have the dimension of an energy
\begin{eqnarray*}
   H_\mathrm{eff}\vert\varphi\rangle = E\vert\varphi\rangle 
\;,\end{eqnarray*}
achieving this way step 3 of Eq.(\ref{eq:1}).
Note that mass and energy in the front form, on the light cone, 
are related by
\begin{equation}
   \hspace{13ex} M^2 = \left(m_1+m_2\right)^2 + 2\left(m_1+m_2\right) E
\;,\label{eq:22}\end{equation}
and \textbf{not} by 
$\;M^2=\left(m_1+m_2\right)^2 + 2\left(m_1+m_2\right) E+E^2$,
as usual. Only if the energy is negligible as compared to the
quark masses, \textit{i.e.} only if $\left(E/(m_1+m_2)\right)^2\ll 1$,
the two relations coincide.

A rather drastic technical simplification is achieved by
a transformation of the integration variable.
One can substitute the integration variable $x$
by the integration variable $k_z$, which, 
for all practical purposes,  
can be interpreted \cite{BroPauPin98} 
as the $z$-component of a 3-momentum vector
$\vec p=(k_z,\vec k_{\!\perp})$.
For equal masses $m_1=m_2=m$, 
the transformation is, together with its inverse,
\begin{eqnarray}
   x(k_z) &=& \frac{1}{2} \left[1+\frac{k_z}
     {\sqrt{m^2 + \vec k_{\!\perp}^{\;2}+ k_z^2}}\right]
\;.\label{eq:23}\\
   k_z^2(x) &=& (m^2+\vec k_{\!\perp}^{\,2})
   \ \frac{\left(x-\frac{1}{2}\right)^2}{x(1-x)} 
\;.\end{eqnarray}
Inserting these substitutions into Eq.(\ref{eq:15})
and defining the reduced wave function $\varphi$ by
\begin{eqnarray}
   \psi_{h_1h_2}(x,\vec k _{\!\perp}) &=&  
   \frac{\sqrt{A(k_z,\vec k _{\!\perp})}}{\sqrt{x(1-x)}}
   \varphi_{h_1h_2}(k_z,\vec k _{\!\perp})    
\,,\label{eq:25}\\
   A(\vec p)&=&\sqrt{1+\frac{\vec p^{\,2}}{m^2}} 
\;,\end{eqnarray}
leads to an integral equation in the components of $\vec p$,
in which all reference to light-cone variables has disappeared.
Using in addition the ST-model of Eq.(\ref{eq:18}),
Eq.(\ref{eq:15}) translates for singlets identically into
\begin{eqnarray}
\lefteqn{ 
   M^2\varphi(\vec p) =
   4\left[m^2+\vec p^2\right] \varphi(\vec p) 
}\label{eq:27}\\ 
   &-& \frac{\alpha_c}{2\pi^2}\int\!\frac{d^3p'}{\sqrt{A(p)A(p')}}
   \left(\frac{4m^2}{Q^2} + 2 \right)
   \frac{R(Q)}{m}\;\varphi(\vec p')
\;,\nonumber\end{eqnarray}
with $\alpha_c=\frac43\alpha_s$.
The equation for the triplets is obtained by dropping the `2'.
In the ST-model, the helicity arguments in the wave functions can be 
suppressed.
Applying the relation between mass and energy, as given in Eq.(\ref{eq:22}),
the equation is converted to
\begin{eqnarray}
\lefteqn{ 
   E\varphi(\vec p) =
   \frac{\vec p ^2}{2m_r}\varphi(\vec p) 
}\label{eq:28}\\ 
   &-& \frac{\alpha_c}{2\pi^2}\int\!\frac{d^3p'}{\sqrt{A(p)A(p')}}
   \left(\frac{4m^2}{Q^2} + 2 \right)
   \frac{R(Q)}{4m^2}\;\varphi(\vec p')
\;,\nonumber\end{eqnarray}
since the reduced mass for $m_1=m_2=m$ is  $m_r=m/2$.

The first term in this equation, $\vec p ^2/2m_r$, coincides with 
the kinetic energy in a conventional non-relativistic Hamiltonian.
This is remarkable in view of the fact that no approximation 
to this extent has been made. 
The fully relativistic and covariant light-cone approach
has no relativistic corrections in the kinetic energy!

Since the first term in Eq.(\ref{eq:28}) is a kinetic energy,
the second must be a potential energy --- in a momentum representation.
In principle, it could be Fourier transformed
with $\mathrm{e}^{-i \vec p \vec r}$ to a configuration space
with the variable $\vec r$. 
But due to the factor $A(p)A(p')$ in the kernel,
the resulting potential energy would be non-local, see 
f.e. \cite{PauMer97}.

The non-locality of the potential is certainly mathematically exact.
But I do not expect this to generate aspects of leading importance, 
and avoid it by the simplification $A(p)\equiv 1$, 
both in Eqs.(\ref{eq:25}) and (\ref{eq:28}).

With $A(p)= 1$,
the mean four momentum transfer $Q^2$ reduces 
to the three momentum transfer $q^2=(\vec p-\vec p')^2$. 
In consequence, the kernel of Eq.(\ref{eq:28}),
\begin{eqnarray}
    U(\vec q) &=&  -\frac{\alpha}{2\pi^2} 
   \left(\frac{4m^2}{q^2}+2\right)
   \frac{R(q)}{4m^2}
\;,\label{eq:29}\end{eqnarray}
depends only on $\vec q=\vec p-\vec p'$.
Its Fourier transform is a local function,  
\begin{eqnarray}
   V(\vec r) &=& \int\!d^3 q\;\mathrm{e}^{-i\vec q \vec r}\;U(\vec q)
\;,\label{eq:30}\end{eqnarray}
which plays the role of a conventional \emph{potential energy} 
in the Fourier transform of Eq.(\ref{eq:28}), \textit{i.e.} in
\begin{eqnarray}
   E\;\psi(\vec r) &=& 
   \left[\frac{\vec p ^2}{2m_r}+V(\vec r)\right]\psi(\vec r) 
\;.\label{eq:31}\end{eqnarray}
Here is the Schr\"odinger equation from Eq.(\ref{eq:1})\,!
Despite its conventional structure it is a front form equation, 
designed to calculate the light-cone wave function 
$\psi(\vec r)\rightarrow\varphi(\vec p)
\rightarrow\psi_{q\bar q}(x,\vec k_{\!\perp})$.

I conclude this section with a subtle point, which 
needs clarification in the future.
The \emph{simplification} $A(p)= 1$ is different from
a \emph{non-relativistic approximation}. 
The approach is certainly valid
also for relativistic momenta $p^2\gg m^2$, particularly
Eqs.(\ref{eq:28}) and (\ref{eq:30}). 
The reason is that $A(p)$ occurs only under the integral. 
There, the large momenta are suppressed by the regulator, anyway. 

\section{The renormalized Coulomb potential}
\label{sec:6}
Hence forward, I restrict consideration to the triplet case,
\textit{i.e.} to Coulomb kernels like 
$U(\vec q) \sim R(q)/q^2 $.
The renormalized Coulomb potential is \emph{always finite at the origin},
as opposed to the conventional $\frac1r$--singularity.
It is instructive to verify this explicitly
for two regulators: 
\begin{eqnarray} 
   U(q) =\left\{
   \begin{array}{ll} 
    -\frac{\alpha_c}{2\pi^2 q^2}\;\frac{\lambda^2}{q^2+\lambda^2},
    &\mbox{for the soft cut-off},\\  
    -\frac{\alpha_c}{2\pi^2 q^2}\;\Theta(q^2-\lambda^2),
    &\mbox{for the sharp cut-off}. 
    \end{array}\right.
\end{eqnarray}
The Fourier transform according to Eq.(\ref{eq:29}) gives
\begin{eqnarray} 
   V(r) =\left\{
   \begin{array}{ll} 
    -\frac{\alpha_c}{r}\;(1-\mathrm{e}^{-\lambda r}),
    &\mbox{for the soft cut-off},\\  
    -\frac{\alpha_c}{r}\;\frac{2}{\pi}\mathrm{Si}(\lambda r),
    &\mbox{for the sharp cut-off}, 
    \end{array}\right.
\end{eqnarray}
where $\mathrm{Si}$ is the Integral Sine.
Asymptotically holds:
\begin{eqnarray} 
   \lim_{r\rightarrow\infty} V(r) =\left\{
   \begin{array}{ll} 
    -\frac{\alpha_c}{r},
    &\mbox{for the soft cut-off},\\  
    -\frac{\alpha_c}{r},
    &\mbox{for the sharp cut-off}. 
    \end{array}\right.
\label{eq:33}\end{eqnarray}
Both cut-offs produce the conventional Coulomb potential.
Near the origin, however, holds:
\begin{eqnarray}
   \lim_{r\rightarrow 0} V(r) &=& \alpha_c\lambda\cdot\left\{
   \begin{array}{ll} 
     -1 + \frac{(\lambda r)\phantom{^2}}{2},
     &\mbox{for the soft cut-off},\\  
     -\frac{2}{\pi}+ \frac{(\lambda r)^2}{9 \pi},
     &\mbox{for the sharp cut-off}. 
    \end{array}\right.
\end{eqnarray}
The renormalized Coulomb potential is \emph{finite} 
but the constant is cut-off dependent. 
Even the $r$-dependence differs:
The soft cut-off gives a linear  
and the sharp cut-off a quadratic dependence.

The cut-off dependence near the origin  
is one of the most important aspects of the present work and 
has a deep physical reason to be discussed below.
Recalling the discussion in Sec.~\ref{sec:3} and
replacing the soft cut-off in analogy to Eq.(\ref{eq:13}) with
\begin{eqnarray}
   R(q) = \left[1+\sum_{n=1}^{N}
   (-1)^n s_n \lambda^n\frac{\partial^n}{\partial\lambda^n}\right]
   \frac{\lambda^2}{\lambda^2+q^2} 
\;,\label{eq:35}\end{eqnarray}
gives straightforwardly the generalized Coulomb potential
\begin{eqnarray}
   V(r) &=& -\frac{\alpha_c}{r} \Big[1+\sum_{n=1}^{N}
   (-1)^n s_n \lambda^n\frac{\partial^n}{\partial\lambda^n}\Big]
   \Big(1-\mathrm{e}^{-\lambda r}\Big) 
\nonumber\\ &=& \phantom{-}
   \frac{\alpha_c}{r}\Big[
   -1+ \mathrm{e}^{-{\lambda r}}\sum_{n=0}^{N}s_n({r\lambda})^n \Big]   ,
\label{eq:36}\end{eqnarray}
with $s_0\equiv1$.
This result illustrates an other important point:
The Laguerre polynomials are a complete set of functions.
The term added to the -1 in Eq.(\ref{eq:36}) is thus potentially
able to reproduce an arbitrary function of $r$. 
The description in terms of a generating function, 
as in Eqs.(\ref{eq:35}) or (\ref{eq:36}), 
is therefore \emph{complete}.

%
%%%%%%%%%%%%%%%%%%%%%%%%%%%%%%%%%%%%%%%%%%%%%%%%%%%%%%%%%%%%%% beg figure
\begin{figure}
\resizebox{0.48\textwidth}{!}{\includegraphics{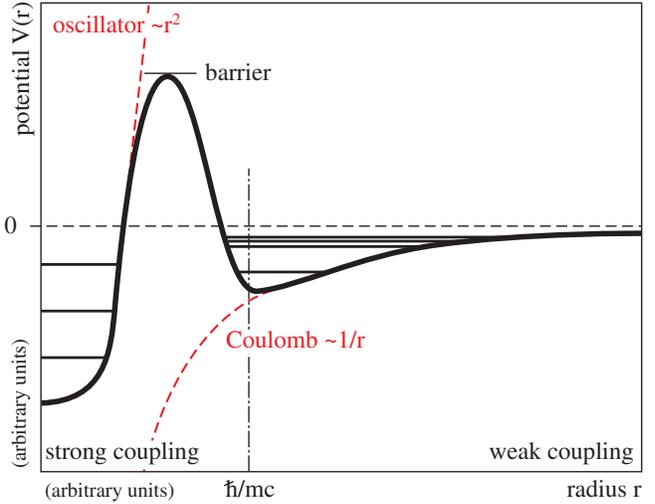}
}\caption{ 
   Schematic behavior of the renormalized Coulomb potential,
   see also the discussion in the text.
}\label{fig:V(r)schem}\end{figure}
%%%%%%%%%%%%%%%%%%%%%%%%%%%%%%%%%%%%%%%%%%%%%%%%%%%%%%%%%%%%%% end figure
%
\textbf{The physical picture} which develops is illustrated 
in Fig.~\ref{fig:V(r)schem}.
In the far zone, for sufficiently large $r$,
the potential energy coincides with the conventional Coulomb potential
$-\frac{\alpha_c}{r}$. 
Since the potential is attractive, it can host bound states
which are probably those realized in weak binding.
In the near zone, for sufficiently small $r$,
the potential behaves like a \emph{power series} $c_0+c_1r+c_2r^2$
which potentially can host the bound states of strong coupling,
provided the actual parameter values allow for that.
In the intermediate zone, the actual potential must interpolate
between these two extremes, since Eq.(\ref{eq:36}) is an analytic
function of $r$. Most likely this is done by developing 
a \emph{barrier of finite} height, depending on the actual parameter
values.
The onset of the near and intermediate regimes must occur
for relative distances of the quarks, which are comparable
to the Compton wave length associated with their reduced mass.
If the distance is smaller, one expects deviations from the classical 
regime by elementary considerations on quantum mechanics, indeed.

%
%%%%%%%%%%%%%%%%%%%%%%%%%%%%%%%%%%%%%%%%%%%%%%%%%%%%%%%%%%%%%% beg figure
\begin{figure} 
   \resizebox{0.48\textwidth}{!}{\includegraphics{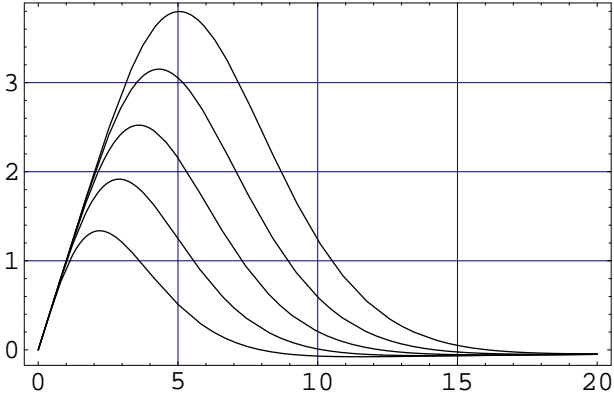}
}\caption{ 
   The dimensionless Coulomb potential $W_N(y;0,1,0)$ is plotted versus 
   the radius parameter $y=\lambda r$ for different $N$, 
   \textit{i.e.} from bottom to top for $N=4,5,6,7,8$. 
}\label{fig:lin}\end{figure}
%%%%%%%%%%%%%%%%%%%%%%%%%%%%%%%%%%%%%%%%%%%%%%%%%%%%%%%%%%%%%% end figure
%
\textbf{The large number of parameter} in Eq.(\ref{eq:36}) 
can be controlled by the following construction:
The coefficients $s_n$ in Eq.(\ref{eq:36})
are expressed in terms of only three parameters $a$, $b$, and $c$, by 
\begin{eqnarray}
    s_n = 
    \frac{1}{n!} + \frac{a}{(n-1)!} + \frac{b}{(n-2)!} + \frac{c}{(n-3)!} 
\,.\label{eq:37}\end{eqnarray}
The first few coefficients are then explicitly
\begin{eqnarray}
   \begin{array}{rc rc rc rc rc rc }
    s_0 &=& 1    &,& \\ 
    s_1 &=& 1 &+&  a &,& \\ 
    s_2 &=& \frac{1}{2}    &+& a &+& b &,& \\  
    s_3 &=& \frac{1}{6}    &+& \frac{a}{2}    &+& b            &+&  c &,& \\ 
    s_4 &=& \frac{1}{24}   &+& \frac{a}{6}    &+& \frac{b}{2}  &+&  c &.&   
   \end{array} 
\end{eqnarray}
As a consequence, the dimensionless Coulomb potential,
\begin{eqnarray}
   W_N(y;a,b,c) = 
   \frac{V(r)}{\alpha_c\lambda} =
   \frac{1}{y}\Big(-1+ \mathrm{e}^{-y}\sum_{n=0}^{N}s_ny^n\Big) 
\,,\label{eq:39}\end{eqnarray}
which depends on $r$ only through the dimensionless combination $y=\lambda r$,
is at most a quadratic function of $y$, 
\begin{eqnarray}
   W_N(y;a,b,c) = a + by + cy^2
\,,\label{eq:40}\end{eqnarray}
in the near zone, and thus independent of $N$. 
The remainder starts at most with power $y^{N+1}$.
A value of $a=c=0$ and $b=1$ should therefore
yield a linear set of functions $W_N(y;a,b,c)=y$
in the near zone. As shown in Fig.~\ref{fig:lin} this happens
to be true for surprisingly large values of $y$,
\textit{i.e.} not only for $y\ll 1$. 
The value of $N$ essentially controls the height of the barrier.
Similarly, $W_N(y;0,0,1)=y^2$ generates a set of functions
which are strictly quadratic in the near zone. 
Again, $N$ controls the height of the barrier, as to be seen below in  
Fig.~\ref{fig:V(r)rho}.
\section{Determining the parameters by experiment}
\label{sec:7}
The QCD-inspired model developed thus far has a considerable
number of renormalization group invariant parameters,
which must be determined once and for all by experiment.

\begin{figure}
   \vspace{-7ex}
   \resizebox{0.48\textwidth}{!}{\includegraphics{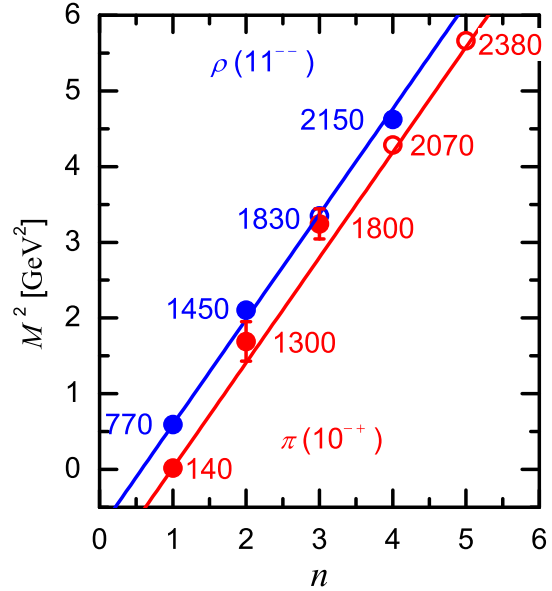}
}\vspace{-4ex}
   \caption{ 
   The invariant mass-squares of all available $\pi^+$-- and 
   $\rho^+$--states are plotted versus a counting index $n=0,1,2,\dots$. 
   The straight lines correspond to $M_n^2=M_0^2+n\chi$, with the value 
   $\chi=1.39\mbox{ GeV}^2$,
   taken from Anisovich \textit{et al.} \protect{\cite{AniAniSar02}}.
   The filled circles correspond to states which have been seen empirically
   \protect{\cite{PDG98}}, the empty ones %circles 
   correspond to the predictions \protect{\cite{AniAniSar02}}.~---
   Plot courtesy of Shan-Gui Zhou.
}\label{fig:AniZhou}\end{figure}

In doing this \cite{FrePauZho02b}, %\cite{FrePauZho02a,FrePauZho02b}, 
we have been inspired by the work of Anisovich \textit{et al.} \cite{AniAniSar02}.
Enumerating the excited states of a hadron by a counting index 
$n=0,1,2,\dots$, these authors have found the linear relation
$M_n^2=M_0^2+n\chi$ for practically all hadrons.
As an example, I present in Fig.~\ref{fig:AniZhou}
the spectrum of the $\pi$- and the $\rho$-meson.
%
%%%%%%%%%%%%%%%%%%%%%%%%%%%%%%%%%%%%%%%%%%%%%%%%%%%%%%%%%%%%%% beg figure
%%%%%%%%%%%%%%%%%%%%%%%%%%%%%%%%%%%%%%%%%%%%%%%%%%%%%%%%%%%%%% end figure
%

The linear relation between mass--squared and energy on the light cone,
Eq.(\ref{eq:22}), allows then to conclude that 
the potential energy in the near zone must be a pure oscillator,
\begin{eqnarray}
    V(r) &=& -c_t + \frac12 f_t r^2 
\;,\label{eq:41}\end{eqnarray}
at least to first approximation, and that $b=0$ in Eq.(\ref{eq:40}).
If one addresses to reproduce the spectra of all flavor off-diagonal 
triplet mesons (pseudo-vector mesons), except the topped ones, 
one has to determine 6 parameters: 
The 2 constants from the oscillator model, $c_t$ and $f_t$, 
and the 4 effective flavor quark masses $m_u=m_d$, $m_s$, $m_c$, 
and $m_b$. To determine them, one needs 6 experimental numbers,
and I take from \cite{PDG98}: 
\begin{eqnarray} 
   \begin{array}{r@{}c@{}l r@{}c@{}l r@{}c@{}l}
     M_{u\bar d,t0} &=& 0.768, & M_{u\bar s,t0} &=& 0.892, & 
     M_{u\bar c,t0} &=& 2.010, \\
     M_{u\bar d,t1} &=& 1.465, & M_{u\bar s,t1} &=& 1.680, &
     M_{u\bar b,t0} &=& 5.325, 
   \end{array} 
\label{eq:42}\end{eqnarray}
all in GeV. The notation should be self-explanatory.
For example, $M_{u\bar d,t1}$ refers to the first excited state of the
$\rho^+$. The so obtained parameter values are:
\begin{eqnarray}
  \begin{array} {|@{\ }c@{\ }|@{\ }c@{\ }|@{\ }c@{\ }|@{\ }c@{\ }%
                ||@{\ }c@{\ }|@{\ }c|}\hline
        m_u   &  m_s   &  m_c  &   m_b  & c_t   & f_t\\ 
       0.218  & 0.438  & 1.749 & 5.068  & 0.880 & 0.0869\\  
     \mathrm{GeV}&\mathrm{GeV}&\mathrm{GeV}&\mathrm{GeV}&
     \mathrm{GeV}&\mathrm{GeV}^3 \\
   \hline
   \end{array} 
\label{eq:43}\end{eqnarray}
The numbers differ slightly from those in \cite{FrePauZho02b},
due to choosing the empirical data set different from Eq.(\ref{eq:42}),
but yield about the same overall agreement with all
available experimental states of pseudo-vector mesons.

Reverting the argument, one concludes as in \cite{FrePauZho02b} 
that the oscillator model in Eq.(\ref{eq:41}) explains 
quite naturally the systematics
found by Anisovich \textit{et al.} \cite{AniAniSar02}.
But one can do even better.

%
%%%%%%%%%%%%%%%%%%%%%%%%%%%%%%%%%%%%%%%%%%%%%%%%%%%%%%%%%%%%%% beg figure
\begin{figure}
   \resizebox{0.48\textwidth}{!}{\includegraphics{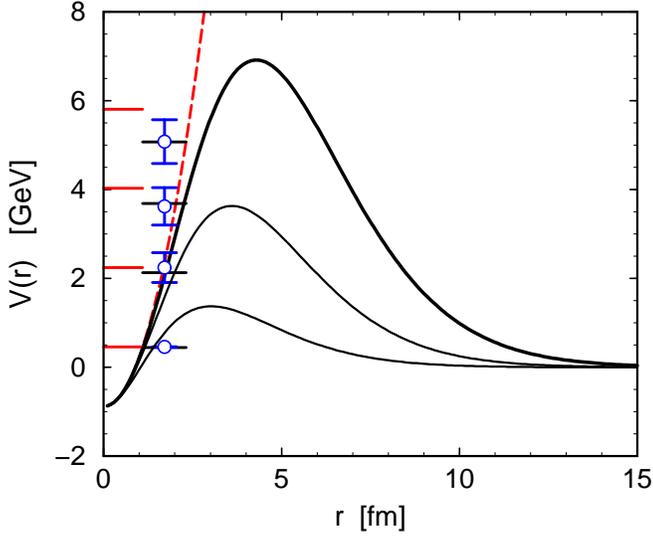}
}\caption{ 
   The continuous lines display the generalized Coulomb potential 
   $V(r)= \alpha_c \lambda W_N (\lambda r;a,0,c) $ 
   in physical units as function of $r$, for the values $N=4,5,6$
   from bottom to top. 
   The circles indicate the experimental eigenvalues $E_n$ 
   for the $\rho^+$. 
   They agree with the calculated eigenvalues for $N=6$, 
   shown by the horizontal lines.~--- 
   The dashed line displays the harmonic approximation; 
   the horizontal lines on the left indicate the oscillator states.~--- 
   See the discussion in the text.
}\label{fig:V(r)rho}\end{figure}
%%%%%%%%%%%%%%%%%%%%%%%%%%%%%%%%%%%%%%%%%%%%%%%%%%%%%%%%%%%%%% end figure
%
%
\section{Relating the oscillator model to QCD}
The oscillator model in Eq.(\ref{eq:41}) is only the harmonic
approximation to the QCD--inspired, 
generalized Coulomb potential in Eq.(\ref{eq:36}).
Their parameters are related obviously by
\begin{eqnarray}
   c_t &=& - \alpha_c\lambda a\,,\qquad
   b = 0 \,,\qquad
   f_t =  2\alpha_c\lambda^3 c
\;.\label{eq:44}\end{eqnarray}
One needs more experimental information to pin down the 
value of $a$, $c$ and $N$. Choosing $\lambda$ as the QCD scale,
\textit{i.e.}
\begin{eqnarray}
   \lambda &=& 200\mbox{ MeV}
\;,\label{eq:45}\end{eqnarray}
one can use the expressions for $\overline \alpha(Q)$ in \cite{Pau98} to 
calculate $\alpha_s\equiv\overline \alpha(0)$ from the
measured value of the coupling constant at the $Z$-mass 
$M_Z=91.2\mbox{ GeV}$, 
\begin{eqnarray}
   \overline \alpha(M_Z) &=& 0.118\,,\quad\mbox{ thus }\quad
   \alpha_s\equiv\overline \alpha(0)=0.1695
\;,\label{eq:46}\end{eqnarray}
as to be shown in greater detail in \cite{Pau03b}.
Having fixed $\alpha_c=\frac43\alpha_s$ and $\lambda$ 
allows to calculate $a$ and $c$ from $c_t$ and $f_t$, \textit{i.e.}
\begin{eqnarray}
   a &=& - 19.5\,,\qquad c = 24.0  
\;.\label{eq:47}\end{eqnarray}
We are thus able to draw the generalized 
Coulomb potential 
$V(r)= \alpha_c \lambda W_N (\lambda r;a,0,c) $
for different $N$ as done in Fig.~\ref{fig:V(r)rho}.
The `experimental' eigenvalues $E_0$---$E_3$
for the $\rho$--meson, obtained 
by means of $E_n = (M_n^2-4m_u^2)/(4m_u)$,
see Eq.(\ref{eq:22}), are also inserted,
including the empirical limits of error. 
The experimental error $\delta E_{\rho,3}\sim \pm 0.5$~GeV 
(thus $\delta M_{\rho,3}\sim \pm 0.1$~GeV) 
is hypothetical, since $M_{\rho,3}$ is not confirmed.
Taking it for granted, the lowest possible value for $N$ is thus
\begin{eqnarray}
   N=6
\;.\label{eq:48}\end{eqnarray}
This completes the determination of all parameters.
They are universal within the model.
I thank Harun Omer\cite{Ome04} for giving me the exact eigenvalues
prior to publication.
\section{Summary and Conclusions}
This work is an important mile stone on the long way 
from the canonical Lagrangian for quantum chromo dynamics 
down to the composition of physical
hadrons in terms of their constituting quarks and gluons,
by the eigenfunctions of a Hamiltonian.

As part of a on-going effort, 
a denumerable number of simplifying assumptions 
had to be phrased for getting a manageable formalism \cite{Pau99b}.
Among them is the formulation of an effective interaction
by the method of iterated resolvents \cite{Pau98},
but the strongest assumption in the present work is probably 
the simplifying Singlet-Triplet model in Sec.~\ref{sec:4}.  
As long as the assumption are not proven at least \textit{a posteriori}, 
one must speak of an approach inspired by QCD.
It is advantageous, however, to have a sufficiently simple formalism
for penetrating the physical content of gauge theory 
by analytical relations.

The biggest progress of the present work can be found in 
Sects.~\ref{sec:2} and \ref{sec:3}.
It is related to a consistent regularization and 
renormalization of a gauge theory.
The ultraviolet divergences in gauge theory are caused less by 
the possibly large momenta of the constituent particles, 
but by the large momentum \emph{transfers} in the interaction.
In a Hamiltonian approach, such as the present, one has not
much choice else than to chop them off 
by a regulating form factor in the elementary vertex interaction. 

The form factor makes its way into a regulator function 
which suppresses the large momentum transfers 
in the Fourier transform of the Coulomb interaction, see Sec.~\ref{sec:6}.
The arbitrariness in chopping off the \emph{large momentum transfers} 
is reflected in the arbitrariness of the potential 
at \emph{small relative distances}.
It is this arbitrariness which allows for a pocket in the 
potential which binds the quarks in a hadron.

The problem is then how to fix this function with its
many parameters, by experiment.
In practice this is less difficult than anti-cipated, see Sec.~\ref{sec:7}.
It suffices to determine only three parameters, two
continuous ones and one counting index.

The potential energy of the present work vanishes
at an infinite separation of the quarks.
This seems be be in conflict with the potential 
energies of phenomenological models \cite{GodIsg85} which rise forever.
It also seems to be in conflict with 
lattice gauge calculations \cite{Schilling2000,Schierholz00}.
Is a finite ionization limit in conflict also with `confinement', 
\textit{i.e.} with the empirical fact that 
free quarks have not been observed?~---
The present model prohibits free quarks as a stable solution,  
since the sum of the constituent quark masses 
is always larger than the mass of the corresponding hadron and a pion.
Free constituent quarks would hadronize very quickly into bound states.
This is different from atomic physics with its free constituents, 
where the binding energy is always much smaller than 
the mass of positronium proper.

The most disturbing aspect of the present work
is its obvious conflict with lattice gauge calculations 
\cite{Schilling2000,Schierholz00} 
and their successes. Several points however should be made:
I have not checked to which extent a linear term
in the potential is consistent with the excellent agreement
between theory and experiment presented in this work.~--
Even with present day computers lattice gauge calculations
can be extrapolated down to such light systems 
as the $\pi$ or the $\rho$ only with a head ake.~--
The calculation of the potential energy on the lattice 
rests on the assumptions of static quarks,
of quarks with an infinitely large mass.
Whether this object is the potential energy 
to be used in a non relativistic Hamiltonian is an open question, 
as well as whether its eigenvalue can simply be added
to the constituent masses to get the invariant
mass of physical hadrons.
In principle, the relation is justified only only 
for sufficiently small coupling constants.

The present work opens a broad avenue of further applications,
among them also the baryons and physical nuclei.
But much work must be done in the future before such a simple 
approach as the present must be taken serious.
It is a first step only.

\end{document}